\documentclass[twoside,twocolumn,9pt]{article}
\usepackage{extsizes}
\usepackage[super,sort&compress,comma]{natbib}
\usepackage[version=3]{mhchem}
\usepackage[left=1.5cm, right=1.5cm, top=1.785cm, bottom=2.0cm]{geometry}
\usepackage{balance}
\usepackage{widetext}
\usepackage{times}
\usepackage{newtxtext,newtxmath}
\usepackage{eqnarray,amsmath}
\usepackage{sectsty}
\usepackage{graphicx}
\usepackage{booktabs}
\usepackage{lastpage}
\usepackage[format=plain,justification=raggedright,singlelinecheck=false,font={stretch=1.125,small,sf},labelfont=bf,labelsep=space]{caption}
\usepackage{float}
\usepackage{fancyhdr}
\usepackage{fnpos}
\usepackage[english]{babel}
\usepackage{array}
\usepackage{droidsans}
\usepackage{charter}
\usepackage[T1]{fontenc}
\usepackage[usenames,dvipsnames]{xcolor}
\usepackage{setspace}
\usepackage[compact]{titlesec}
\usepackage{bm}
\usepackage{multicol}

\usepackage{epstopdf}

\definecolor{cream}{RGB}{222,217,201}

\begin{document}



\makeFNbottom
\makeatletter

\def\red{\textcolor{red}}

\def\Gr{{\rm \Gamma}}
\newcommand{\be}{\begin{equation}}
\newcommand{\ee}{\end{equation}}
\newcommand{\bea}{\begin{eqnarray}}
\newcommand{\eea}{\end{eqnarray}}

\renewcommand\LARGE{\@setfontsize\LARGE{15pt}{17}}
\renewcommand\Large{\@setfontsize\Large{12pt}{14}}
\renewcommand\large{\@setfontsize\large{10pt}{12}}
\renewcommand\footnotesize{\@setfontsize\footnotesize{7pt}{10}}
\makeatother

\renewcommand{\thefootnote}{\fnsymbol{footnote}}
\renewcommand\footnoterule{\vspace*{1pt}%
\color{cream}\hrule width 3.5in height 0.4pt \color{black}\vspace*{5pt}}
\setcounter{secnumdepth}{5}

\makeatletter
\renewcommand\@biblabel[1]{#1}
\renewcommand\@makefntext[1]%
{\noindent\makebox[0pt][r]{\@thefnmark\,}#1}
\makeatother
\renewcommand{\figurename}{\small{Fig.}~}
\sectionfont{\sffamily\Large}
\subsectionfont{\normalsize}
\subsubsectionfont{\bf}
\setstretch{1.125} 
\setlength{\skip\footins}{0.8cm}
\setlength{\footnotesep}{0.25cm}
\setlength{\jot}{10pt}
\titlespacing*{\section}{0pt}{4pt}{4pt}
\titlespacing*{\subsection}{0pt}{15pt}{1pt}

\fancyfoot{}
\fancyfoot[LE]{\footnotesize{\sffamily{\hspace{2pt}\thepage}}}
\fancyfoot[RO]{\footnotesize{\sffamily{\hspace{2pt}\thepage}}}

\fancyhead{}
\renewcommand{\headrulewidth}{0pt}
\renewcommand{\footrulewidth}{0pt}
\setlength{\arrayrulewidth}{1pt}
\setlength{\columnsep}{6.5mm}
\setlength\bibsep{1pt}

\makeatletter
\newlength{\figrulesep}
\setlength{\figrulesep}{0.5\textfloatsep}

\newcommand{\topfigrule}{\vspace*{-1pt}%
\noindent{\color{cream}\rule[-\figrulesep]{\columnwidth}{1.5pt}} }

\newcommand{\botfigrule}{\vspace*{-2pt}%
\noindent{\color{cream}\rule[\figrulesep]{\columnwidth}{1.5pt}} }

\newcommand{\dblfigrule}{\vspace*{-1pt}%
\noindent{\color{cream}\rule[-\figrulesep]{\textwidth}{1.5pt}} }

\makeatother

\twocolumn[
  \begin{@twocolumnfalse}
\sffamily
\begin{tabular}{m{0.0cm} p{16.0cm} }

\quad &
\noindent\LARGE{\textbf{Modeling of polymer-enzyme conjugates formation: Thermodynamic perturbation theory and computer simulations}} \\
\vspace{0.3cm} & \vspace{0.3cm} \\
 & \noindent\large{Halyna Butovych, Yurij V. Kalyuzhnyi, Taras Patsahan, Jaroslav Ilnytskyi} \\
 \quad & \textit{Institute for Condensed Matter Physics of the National Academy of Sciences of Ukraine} \\ 
 &\textit{1 Svientsitskii St., Lviv, Ukraine 79011; E-mail: tarpa@icmp.lviv.ua} \\ 
 \vspace{0.3cm} & \vspace{0.3cm} \\
\quad 
& \noindent\normalsize
{A simple model for the formation of the polymer-enzyme conjugates has been proposed and
described using corresponding extension of the Wertheim's first-order thermodynamic 
perturbation theory (TPT1) for the system of associating chain molecules. 
A set of computer simulation data for different number of functional groups along polymer chains has been obtained and used to access the accuracy of the theoretical results. Predictions of the present theoretical approach are more accurate than that of the conventional TPT1 and are in a very good agreement with the computer simulation data. In particular the theory is able to account for the difference in position of the polymer functional groups along its backbone.}
\end{tabular}

 \end{@twocolumnfalse} \vspace{0.6cm}

]

\renewcommand*\rmdefault{bch}\normalfont\upshape
\rmfamily
\section*{}
\vspace{-1cm}


\section{Introduction}

The application of enzymes in chemical industry is of great interest and it is constantly developing. This is primarily because of unique catalytic properties of enzymes, which allow them to be considered as an environmentally friendly alternative to many traditional toxic and hazardous technologies. One of the areas of such application is converting cellulose biomass into biofuel. The enzymatic biocatalysis
is pivotal in the production of first and second generation bioethanol from cellulose using a mixture of different types of enzymes, so-called cellulases (endocellulases, cellobiohydrolases, beta-glucosidases)~\cite{Wyman2011}. Such kind of enzymes are involved in the process of cellulose hydrolysis, where they synergistically break polysaccharide chains into monosaccharide and oligosaccharide molecules, from which ethanol is subsequently produced by fermentation. Despite the attractiveness of this approach, there is still the problem of efficiency and commercial viability of this technology~\cite{Bao2016}. There are a number of strategies to increase a catalytic efficiency of enzymes~\cite{Chinn2015}. One of the most promising approaches is immobilization of enzymes on scaffolds~\cite{Grove2016,Han2016,Bayer2010,Bayer2016,Li2021} like polymers~\cite{Minko2021}, nanoparticles\cite{Bordbar2017} and microgels. For instance, it was found the synthetic enzyme-polymer conjugates mimicking natural cellulosomes can exhibit a sufficiently higher catalytic activity in cellulose hydrolysis if to compare with dispersions of totally free enzymes~\cite{Han2016,Bayer2010,Bayer2016,Minko2021}. This obviously indicates a higher synergism level of enzymes constrained into groups. Moreover, the enzymes captured on scaffolds can be more easily collected for
further reutilization. It was also observed that enzymes in complexes with brush-like polymers have a higher thermal stability~\cite{Minko2017,Minko2018}.

\begin{figure}[!h]
	\centering
	\includegraphics[clip,height=0.2\textwidth,angle=0]{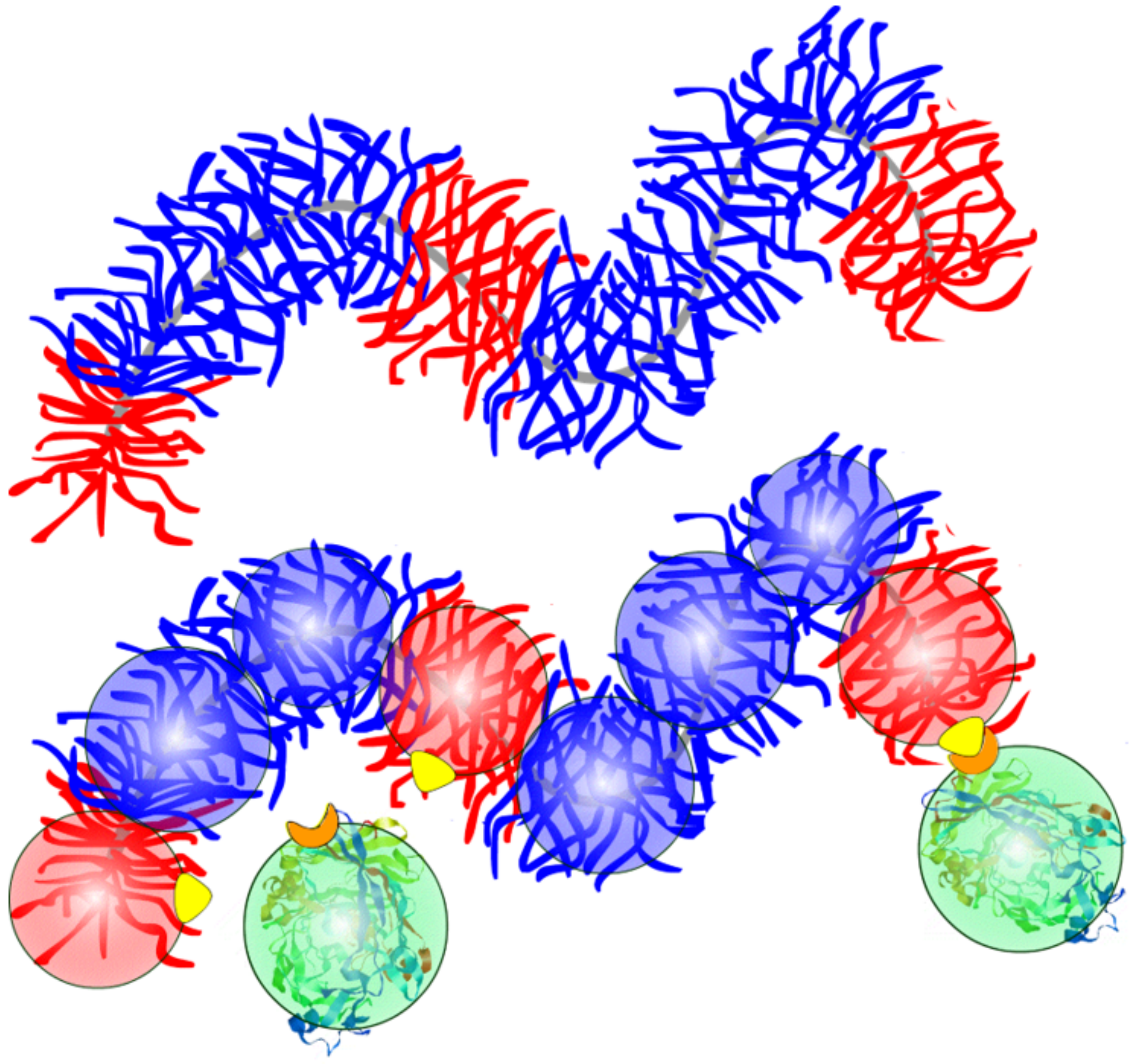} 
\caption{\label{fig:brush12} Sketch of molecular brush block copolymer (top) and its coarse-grained representation with enzyme molecules (bottom).}
\end{figure}

We report a theoretical study of conjugation process in a mixture of enzymes with brush block copolymers. We propose the very basic coarse-grained model 
to describe a mixture of enzyme globules and multifunctional polymer chains. Brush copolymer is mimicked by linear chain of spherical beads with an implicit side-chains representation. It is assumed that some of polymer blocks can bear functional groups specific to active sites located at enzyme molecule, which can bind to polymer by affinity interactions~\cite{Han2016,Beloqui2020}.
The simplicity of considered model is motivated by two reasons. The first is to distinguish the role of functional groups arrangement
at polymer chains playing in the polymer-enzyme conjugation. Secondly, we are aimed to develop a theoretical approach to describe this process. 
In fact, it still remains a challenge to provide a fast and sufficiently accurate theoretical prediction for properties 
of polymer-protein mixtures made of beads bearing attractive sites even for simple models. For this purpose, some approaches of the associative liquids theory seems to be the most appropriate ones~\cite{Wertheim1984,Wertheim1986,Wertheim1987}, however they need an essential improvement.

In this study we focus on the development of the model and its theoretical description of the polymer-enzyme conjugation process. With
this goal in mind we extend and apply the first order multi-density thermodynamic perturbation 
theory (TPT1) of Wertheim~\cite{Wertheim1984,Wertheim1986,Wertheim1987}
and perform detailed computer simulation study of the model. Our theoretical and computer
simulation results are compared and the accuracy of the theoretical predictions are accessed.
The models similar to that we suggested have been used earlier to describe effects of aggregation and 
association in the surfactant systems, alkanols, organic acids, aminoacids, etc (see~\cite{Muller2001,Economou2002,Paricaud2002,Tan2008,McCabe2010} 
and references therein). 
In these studies the particles of the system were modeled by the associating
totally flexible chains of spherical Lennard-Jones (LJ) monomers (beads). Similar as in our model 
association occurs due to the off-center short-ranged square-well bonding sites placed on the surface of 
certain monomers. For the theoretical description the authors have been using TPT1 with the reference system 
represented by the fluid of the LJ monomers. The theory, being relatively successful in describing the 
models with bonding sites placed on the terminal beads of the chains, appears much less accurate if
bonding sites were located on the intermediate beads~\cite{Muller1994,Muller1995,Herdes2004}. 
This drawback of the theory is due to the "single bond" approximation utilized in the TPT1, i.e. bonding abilities 
of each of the sites of the monomer do not depend on the bonding state of the rest of the sites~\cite{Wertheim1984,Wertheim1986,Wertheim1987}. 
There are two possibilities to improve performance of the theory: to apply the higher-order approximations 
in the framework of the TPT or to use the better choice for the reference system. 
Unfortunately, the former option is not much feasible, since it requires
the application of the higher-order reference system correlation functions, which restrict practical application of this scheme 
for the second order case~\cite{Wertheim1987}. To achieve the higher accuracy using low-order version of the TPT 
in many cases it is more profitable to stick with the former option. 
This route was undertaken in~\cite{Kalyuzhnyi1994,Rescic2016,Kalyuzhnyi2017a,Kalyuzhnyi2017b} to account for the effects of changes 
in the system excluded volume upon association. In this study we propose to use the reference system 
represented by the fluid of nonassociating chain molecules. Somewhat similar approach was used to correct
TPT1 predictions for the low-density behavior of the equation of state for the chain fluids~\cite{Ghonasgi1994,Chang1994}. 
However, to the best of our knowledge this idea has not been used for the
description of the associating chains, in particular when association occurs due to the intermediate
monomers of the chain. In addition, in the majority of these studies the properties of the reference 
fluid were calculated using the fit of the corresponding computer simulation results. Here description 
of the reference system is carried out analytically, using solution of the multi-density Ornstein-Zernike (OZ) equation 
supplemented by the associative Percus-Yevick (APY) approximation~\cite{Kalyuzhnyi1995,Kalyuzhnyi1997,Lin1998}.

The paper is organized as follows. First, in Section~\ref{sec:model} we introduce  the general model, which can be described by the developed theory and applied in particular to the problem stated above. Then we present the theory in Section~\ref{sec:theory}. Technical details of computer simulation performed in this study can be found in Section~\ref{sec:simulation}. The obtained results are discussed in Section~\ref{sec:results} and concluding remarks are presented in the last Section~\ref{sec:conclusions}.

\section{The model}\label{sec:model}

We start with the two-component mixture of flexible chain molecules consisting of tangentially bonded hard-sphere monomers of different size. Some of the monomers in the chains of one type and one monomer in 
the chains of the other type bear single off-center attractive square-well site (functional group), located on the surface.
This site-site attractive interaction is valid only between the sites, which belong to the chains of 
different type. 
In the limiting case of the latter component represented by only one functional monomer this model reduces to
the model of polymer-enzyme mixture.
The pair potential acting between the monomers is
\be
U_{a_ib_j}(12)=U_{a_ib_j}^{(hs)}(r)+U_{a_ib_j}^{(sw)}(12),
\label{Potential}
\ee
where $a(b)$ denote the type of the chain and $i(j)$ denote the type of the monomer belonging to the
chain of the type $a(b)$,
\be
U^{(hs)}_{a_ib_j}(r)
=\left\{\begin{array}{cc}
	\infty,&\;\;{\rm for}\;\; r \leq \sigma_{a_ib_j} \\
	0,&\;\;\;\;\;{\rm otherwise}
\end{array} \right.,
\label{hs}
\ee
\be
U^{(sw)}_{a_ib_j}(12)\equiv U^{(sw)}_{a_ib_j}(z)
=\left\{\begin{array}{cc}
	-(1-\delta_{ab})\epsilon_{ij},&\;\;{\rm for}\;\; z\leq \kappa \\
	0,&\;\;\;\;\;{\rm otherwise}
\end{array} \right.,
\label{conical}
\ee
$z$ is the distance between the square-well sites, $\sigma_{a_ib_j}$ is the hard-sphere diameter, 
$\epsilon_{ij}$ and $\kappa$ are the square-well depth and width, respectively. The indices
$a$ and $b$ each take the values 1 and 2, and the indices $i$ and $j$ take the values $1,\ldots,n_a$
and $1,\ldots,n_b$, respectively. Here $n_a(n_b)$ is the number of the monomers in the chain of the 
type $a(b)$. The system number density is $\rho=\rho_1+\rho_2$, where $\rho_1$ and $\rho_2$ are the
number density of the chains of the type $1$ and $2$, respectively. 

\section{Theory}\label{sec:theory}
Following Wertheim \cite{Wertheim1984,Wertheim1986} we assume that Helholtz free energy $A$ of the 
mixture at hand can be written as a sum of Helmholtz free energy of the reference system $A_{ref}$ and
corresponding contribution due to bonding $\Delta A_{bond}$, i.e.
\be
A=A_{ref}+\Delta A_{bond},
\label{free}
\ee
where the reference system is represented by the original two-component mixture of hard-sphere chain molecules with $\epsilon_{ij}=0$ and for $\Delta A_{bond}$ we have
\be
\beta{\Delta A_{bond}\over V}=\sum_{a=1}^2\rho_a\left[\sum_{i=1}^{n_a}\left(\ln{X_{a_i}}
-{1\over 2}X_{a_i}\right)+{1\over 2}n_a\right].
\label{freeb}
\ee
Here $n_a$ is the number of the square-well sites located on the chain molecule of the type $a$
and $X_{a_i}$ is the fraction of the molecules of the type $a$ with nonbonded square-well site 
on the monomer of the type $i$. These fractions satisfy the mass action law type of the relation
\be
X_{a_i}\sum_{b=1}^2\rho_b\sum_{j=1}^{n_b}X_{b_j}K_{a_ib_j}+X_{a_i}-1=0,
\label{mass}
\ee
where 
\be
K_{a_ib_j}=4\pi\int_{\sigma_{a_ib_j}}^{\sigma_{a_ib_j}+\kappa}r^2{\bar f}_{a_ib_j}^{(sw)}(r)
g_{a_ib_j}^{(ref)}(r)dr,
\label{K}
\ee
$g_{a_ib_j}^{(ref)}(r)$ is the monomer-monomer pair distribution function and 
$ {\bar f}_{a_ib_j}^{(sw)}(r)$ is the orientationally averaged monomer-monomer Mayer function
\be
{\bar f}_{a_ib_j}^{(sw)}(r)=
{(1-\delta_{ab})(e^{\beta\epsilon_{ij}}-1)
\over
6\sigma_{a_ib_j}^2r}
\left(\kappa+\sigma_{a_ib_j}-r\right)^2
\left(2\kappa-\sigma_{a_ib_j}+r\right)
,
\label{Mayer}
\ee

\subsection{Description of the reference system}
Thermodynamic properties of the reference system can be calculated using first-order thermodynamic
perturbation theory (TPT1) of Wertheim \cite{Wertheim1986,Wertheim1987}. For Helmholtz free energy of the reference system we have 
\be
\beta{A_{ref}\over V}=-\sum_{a=1}^2\rho_a\sum_{i=1}^{n_a-1}\ln{y^{(hs)}_{a_ia_{i+1}}(\sigma_{a_ia_{i+1}}}),
\label{Aref}
\ee
where $y^{(hs)}_{a_ib_j}(r)$ is the hard-sphere cavity correlation function. The fractions $X_{a_i}$
follow from the solution of equation (ref{mass}) provided that the value of the integral 
$K_{a_ib_j}$ (\ref{K}) is known. To simplify calculation of this integral we assume that the Mayer function
${\bar f}^{(sw)}_{a_ib_j}(r)$ can be approximated by the Dirac delta-function, i.e.
\be
{\bar f}^{(sw)}_{a_ib_j}(r)=T_{a_ib_j}\delta(r-\sigma_{a_ib_j}),
\label{Baxter}
\ee
where
\be
T_{a_ib_j}=\sigma^{-2}_{a_ib_j}\int_{\sigma_{a_ib_j}}^{\sigma_{a_ib_j}+\kappa}r^2{\bar f}_{a_ib_j}^{(sw)}(r)dr.
\label{T}
\ee
Now for $K_{a_ib_j}$ we have 
\be
K_{a_ib_j}=4\pi\sigma_{a_ib_j}^2T_{a_ib_j}g_{a_ib_j}^{(ref)}(\sigma_{a_ib_j}),
\label{Kdelta}
\ee
where the contact values of the radial distribution functions $g_{a_ib_j}^{(ref)}(\sigma_{a_ib_j})$
can be calculated using analytical solution of the polymer Percus-Yevick ideal chain approximation for
heteronuclear hard-sphere chain fluids \cite{Kalyuzhnyi1995,Kalyuzhnyi1997,Lin1998}
\bea
&&\sigma_{a_ib_j}
g^{(ref)}_{a_ib_j}(\sigma_{a_ib_j})={\sigma_{a_ib_j}\over 1-\eta}+{1\over 4}
{\rho s_n\sigma_{a_i}\sigma_{b_j}\over (1-\eta)^2}\\ \nonumber
&&+{1\over 4(\eta-1)}\sigma_{b_i}\left[
   \left(1-\delta_{j1}\right){\sigma_{b_{j-1}}\over\sigma_{b_jb_{j-1}}}+
\left(1-\delta_{jn_{b}}\right){\sigma_{b_{j+1}}\over\sigma_{b_jb_{j+1}}}\right]\\\nonumber
&&+{1\over 4(\eta-1)}\sigma_{a_i}\left[
   \left(1-\delta_{i1}\right){\sigma_{a_{i-1}}\over\sigma_{a_ia_{i-1}}}+
\left(1-\delta_{in_{a}}\right){\sigma_{a_{i+1}}\over\sigma_{a_ia_{i+1}}}\right]\\\nonumber
&&+{\delta_{ab}\over 8\pi\rho_a}\left[
{(1-\delta_{1i})(1-\delta_{2i})\over\sigma_{a_ia_{i-1}}\sigma_{a_{i-1}a_{i-2}}}\delta_{ij+2}+
{(1-\delta_{in_a})(1-\delta_{in_a-1})\over\sigma_{a_ia_{i+1}}\sigma_{a_{i+1}a_{i+2}}}\delta_{ij-2}
\right].
\label{cont}
\eea
Here $\eta=\pi/6\sum_a\rho_a\sum_i\sigma_{a_i}^3$ and $s_n=\pi/\rho\sum_a\rho_a\sum_i\sigma_{a_i}^2$.

\section{Computer simulation details}\label{sec:simulation}
\begin{figure}[!h]
	\centering
	\includegraphics[clip,height=0.25\textwidth,angle=0]{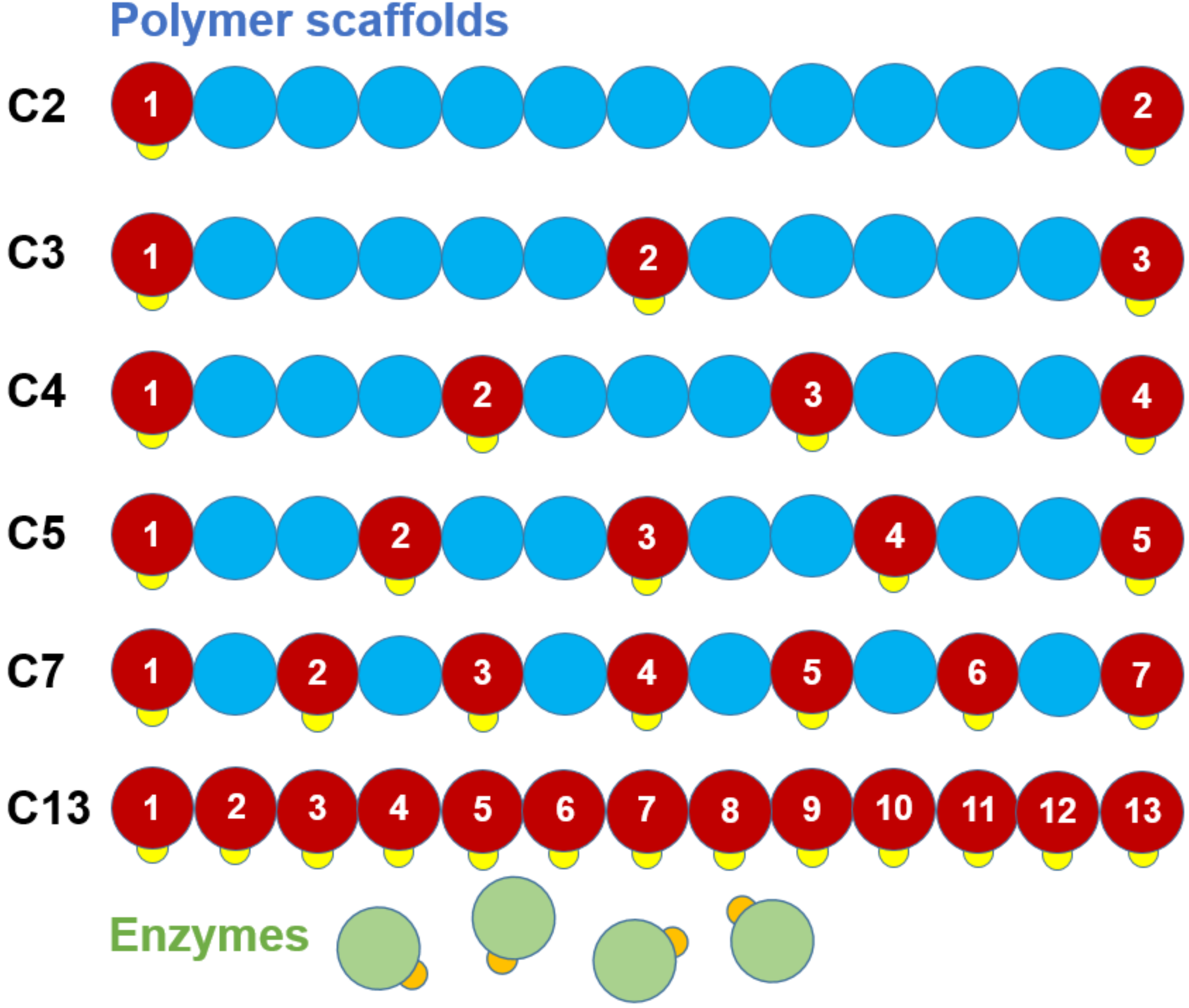}
	\caption{\label{fig:models} The models of brush block copolymer scaffolds and enzymes considered in this study. Red circles denote polymer beads bearing the functional group (yellow spot), blue circles -- generic polymer beads,  green cicrles -- enzyme molecules with the functional group (orange spot).}
\end{figure}

Computer simulations were performed of the models of polymer-enzyme mixtures (see \figurename{\ref{fig:models}}) using the method of Langevin dynamics~(LD) in the NVT ensemble~\cite{Tildesley2017} with the LAMMPS software package (https://www.lammps.org, version 30Nov2020)~\cite{LAMMPS}. Since force fields in LD simulations require continuous pair potentials, the square-well and hard-sphere potentials used in the theory should be substituted with their continuous analogs. 
Therefore, for the square-well interaction~(\ref{conical}) acting between functional sites of polymer and enzyme the original code of LAMMPS was modified by implementing the force field, which according to~\cite{Espinosa2019} describe the continuous square well (CSW) pair potential:
\be
u_{\rm{csw}}(r)=-\frac{1}{2}\epsilon_{\rm{csw}}\left[1-\tanh\left(\frac{(r-r_\text{w})}{\alpha}\right)\right].
\ee
In contrast to \cite{Espinosa2019} we use more steep square well shape of $u_{\rm{csw}}$ by taking $\alpha=0.001\sigma$. The radius of attractive well was chosen $r_{\rm{w}}=0.12\sigma$, the cutoff radius was $0.17\sigma$ and the attractive well depth $\epsilon_{\rm{csw}}$ was defined as the energy unit.
With these parameters the shape of $u_{\rm{csw}}(r)$ very closely fit the conventional square well potential~(\ref{conical}). All functional sites are fixed on the corresponding enzyme and polymer beads at the distance $\sigma/2$ to their centers using the SHAKE algorithm with the tolerance $0.0001$. Sizes of all beads in the polymer and size of enzyme molecules were taken equivalent and set to $\sigma$, which was used as the unit of distances in our simulations.

The hard-sphere pair interaction between particles (\ref{hs}) was substituted with the pseudo-hard sphere pair potential (PHS) using the repulsive part of the cut and shifted (50,49)-Mie potential as it was suggested in~\cite{Jover2012,Espinosa2019}: 
\be
u_{\rm{phs}}(r)
=\left\{\begin{array}{cc}
	50\left(\frac{50}{49}\right)^{49}\epsilon_{\rm{phs}}\left[\left(\frac{\sigma}{r}\right)^{50}-\left(\frac{\sigma}{r}\right)^{49}  \right],& r < \frac{50}{49}\sigma \\
	0,&r \geq \frac{50}{49}\sigma
\end{array} \right.
\label{hs}
\ee
Since we considered two different temperatures in our study, the parameter $\epsilon_{\rm{phs}}$ of PHS potential was rescaled respectively as $\epsilon_{\rm{phs}}=\epsilon_{\rm{csw}} T^{*}/1.5$, where $T^{*}=k_BT/\epsilon_{\rm{csw}}$ was the temperature in reduced units.
The intramolecular interactions within polymer chains were modeled by the harmonic bonds $u_{\rm{bond}}(r)=K(r-r_0)^2$ with the spring constant $K=160.0\epsilon_{\rm{csw}}/\sigma^2$ and the equilibrium distance $r_0=\sigma$. The non-bonding interactions $u_{\rm{csw}}$ and $u_{\rm{phs}}$ were not take into account
between the nearest adjacent beads in the same polymer.

The timestep for the Verlet integration of equations of motion in our simulations was chosen as $\delta t=0.0001\tau$ and the LD damping parameter was set to $\gamma=0.1\tau$, where $\tau=(m \sigma^2\epsilon_{\rm{csw}}^{-1})^{1/2}$ was the time unit. The small timestep guaranteed a stable binding process between polymer and enzyme molecules, which was noticed to be vulnerable at larger timesteps due to the steep pair potentials and very short $r_{\rm{w}}$ used in the model.

All simulations of polymer chains and enzyme molecules were performed in the cubic box of the same size $L=50.0\sigma$ with the periodic boundary conditions applied.
A number of particles varied depending on the number densities of polymer and enzyme molecules taken in our study (see \tablename{\ref{tab:poly_dens}}).
The maximum number of enzymes (about $15000$) was considered in Model~C13. The maximum number of polymer chains ($2000$) was taken in Model~C2, and the minimum of them was in Model~C13 ($1154$).
For each set of parameters the simulations were done in three stages. First stage was used to thoroughly mix enzymes with polymers to lose a trace of the initial system, and it was realized during a short period of $10^6$ simulation steps without patches activated. At the second stage the patches were activated and the system was equilibrated during the period necessary to reach a saturation of bonds creation, which was typically $200\cdot 10^6$ steps. The third stage was for producing results, and it took $300\cdot 10^6$ simulation steps. During the production runs all necessary data were collected at every $10000$ steps and averaged.

\begin{table}[!h]
\caption{\label{tab:poly_dens} Number density of polymer chains taken in each of the models, $\rho_p$.} 
\begin{tabular}{llllll}
\toprule
C2    & C3    & C4    & C5    & C7    & C13   \\ 
\midrule
0.016 & 0.015 & 0.014 & 0.013 & 0.012 & 0.009 \\
\bottomrule
\end{tabular}
\end{table}

\section{Results and discussion} \label{sec:results}

To illustrate application of the TPT1 developed above and to systematically access its accuracy we 
have studied six different grafting densities of the functional groups distributed along polymer chains,
and denote the corresponding systems type as the Models C2, C3, C4, C5, C7 and C13 (see \figurename{\ref{fig:models}). 
The chains representing polymer scaffolds consist of $13$ hard-sphere beads of the size $\sigma$ and 
the number in our notation for the type of the model corresponds to the number of the polymer beads with functional groups. 
These beads are placed along the chain backbone uniformly and symmetrically. 
Enzyme molecules are represented by the hard spheres of the same size $\sigma$ and one functional site which can conjugate to the polymer functional group due to the square-well attractive potential.
The width and depth of the square-well site-site interaction was chosen to be $\kappa=0.119\sigma$ and $\epsilon_{ij}=\epsilon$ 
and we consider two values of the temperature: $T^*=T/\epsilon=0.09$ and $0.12$.

\begin{figure}[!h]
	\centering
	\includegraphics[clip,height=0.3\textwidth,angle=0]{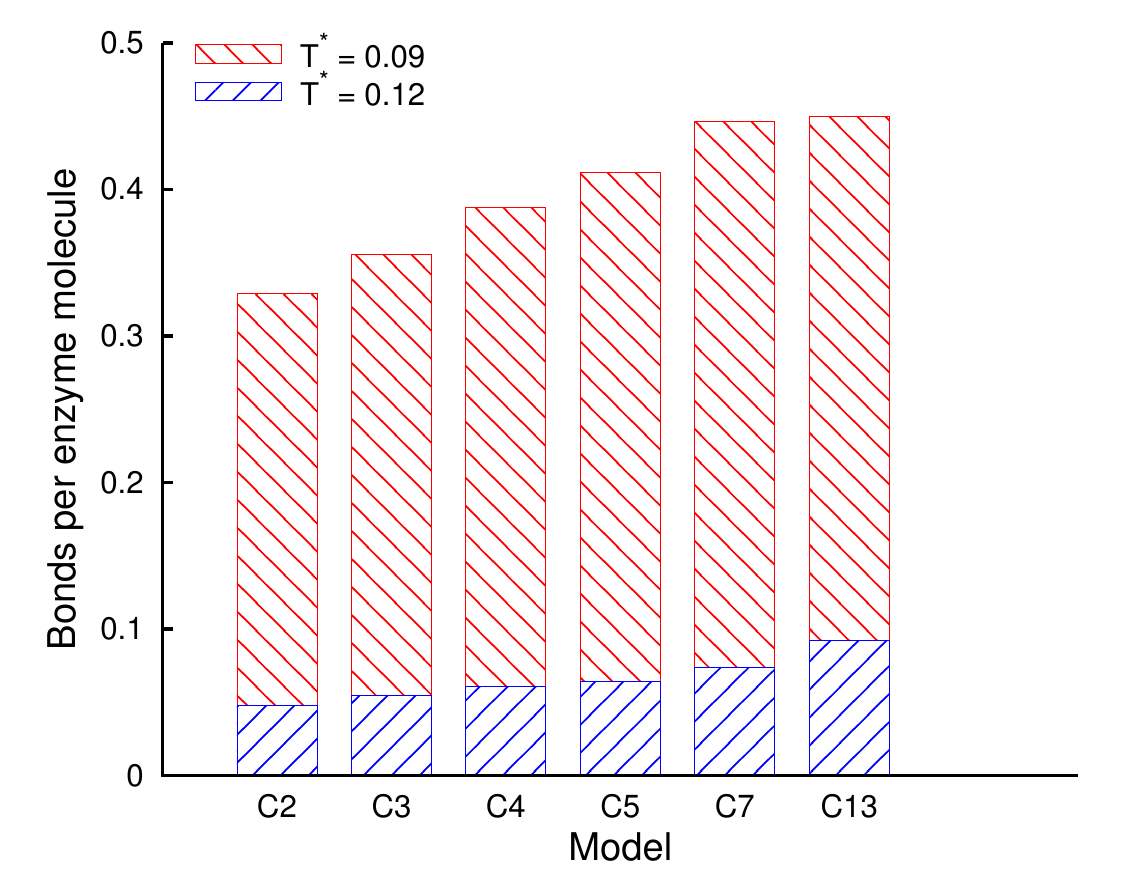} \\
	\includegraphics[clip,height=0.3\textwidth,angle=0]{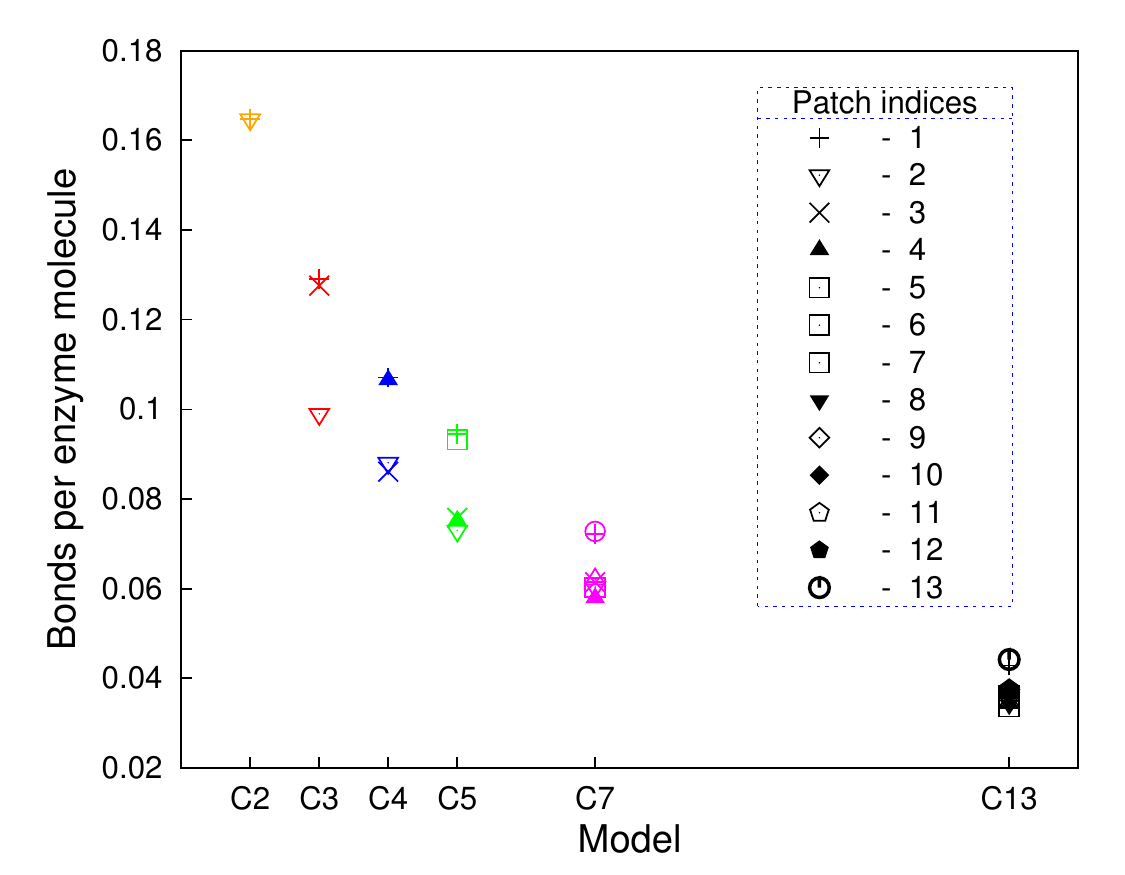}
	\caption{\label{fig:c_frac} Top panel: Average number of bonds per number of enzyme molecules for Models~C2, C3, C4, C5, C7 and C13 at the fixed packing fraction 
	$\eta=0.126$ when the total number of polymer functional groups in the system are equal to the number of enzymes, i.e. $n_e=n_p k_p$ ($k_p$ -- number of patches per polymer chain,
	$n_e$ and $n_p$ -- number of enzyme and polymer molecules, respectively).
	Bottom panel: Number of bonds with the selected polymer functional groups per number of enzyme molecules (for $n_e=n_p k_p$), where each functional group index corresponds to its serial number at polymer molecule ($T^*=0.09$). (see Fig.~\ref{fig:models}). }
\end{figure}

In \figurename{\ref{fig:c_frac}} (top panel) we show the average number of bonds per molecules of enzyme for all six 
types of model at the constant packing fraction of the system  $\eta=\pi/6(\rho_e+13\rho_p)\sigma^3=0.126$ and
at the fixed ratio $\rho_e/\rho_p=k_p$, where $k_p$ is the number of beads of the chain with functional groups.
With this choice of the densities the number of enzyme molecules are equal to the total number of polymer
beads with the functional groups in the system. At higher temperature ($T^*=0.12$) one can observe a slight increase of the fraction 
of bonded enzymes with increasing the number of functional beads per chain at higher temperature. 
At lower temperature this increase is more pronounced, however only up to $7$ functional beads (Model~C7). 
With further a increase of the number of functional beads (Model~C13) the fraction of bonded enzyme remains unchanged. 
Thus, while at the higher $T^*$ the probability of bonding does not depend much on the number of functional beads on
the polymer chain, at the lower temperature it is seen that the large $k_p$ the enzyme molecules begin
to partially block each other upon bonding to nearby beads. 
In \figurename{\ref{fig:c_frac}} (bottom panel) we present the fraction of enzymes bonded to particular functional beads on
the polymer chain. One can observe a substantial difference in the fraction of enzymes bonded
at the terminal beads (chain ends) or to the beads, which are located inside the chain.
The difference between fractions of enzymes bonded to the different beads inside the chain is
minor.

Next we have studied the system at fixed the densities of the chain molecules $\rho_1=\rho_p$ and different densities of the enzymes 
$\rho_2=\rho_e$ in the range of $0-0.18$ and two values of the temperature, $T^*=0.09$, $0.12$. The density of the 
polymer chains were chosen to be different for each model: (see Table 1).
Our theoretical (mTPT1) and computer simulation (symbols) results are shown in \figurename{\ref{fig:c2frac}-\ref{fig:c7frac}}. 
They are presented in reduced units: the distance is measured in the units of $\sigma$ and the energy (temperature) in the units of the square-well depth $\epsilon$. In addition, we include results of the conventional TPT1 approach (TPT1), which is based on the 
application of the hard-sphere reference system. One can observe a very good agreement of our theoretical predictions with the data of computer simulations. 
On the other hand, the theoretical results of the unmodified version of the TPT1 is much less accurate. 

\section{Conclusions}\label{sec:conclusions}
We propose a theoretical description for the formation of 
polymer-enzyme conjugates using a simple model of two-component mixture consisting of flexible hard-sphere chains of polymer bearing a number of uniformly arranged functional groups and hard-sphere enzyme molecules with a single specific site able to conjugate to these groups. 
Depending on the grafting density of functional groups on the polymers and the polymer-enzymes composition, different number of enzymes can bind to the polymer scaffolds at different places.
Theoretical description of the model was carried out using a corresponding extension of the Wertheim's first-order thermodynamic perturbation theory (TPT).
To assess the accuracy of the theory we have performed a set of computer simulation data. 
Our analyses is focused on the abilities of the theory to correctly reproduce 
the number of the bonds created between enzyme molecules and polymer functional groups differently located along the chain. 
We have shown that predictions of the present theory is in a very good
agreement with corresponding computer simulation and appears to be
much more accurate than that of the conventional TPT1 approach. 
An important advantage of the developed theory is its simplicity 
and ability to provide a completely analytical description for the general
case of any number of the functional groups at polymer beads and enzymes molecules as well as for different size ratios of them.

\section*{Acknowledgements}
Yu.K., T.P. and Ja.I. acknowledge CRDF Global Award no. 66705.

\begin{figure*}[!h]
	\centering
	\includegraphics[clip,height=0.35\textwidth,angle=0]{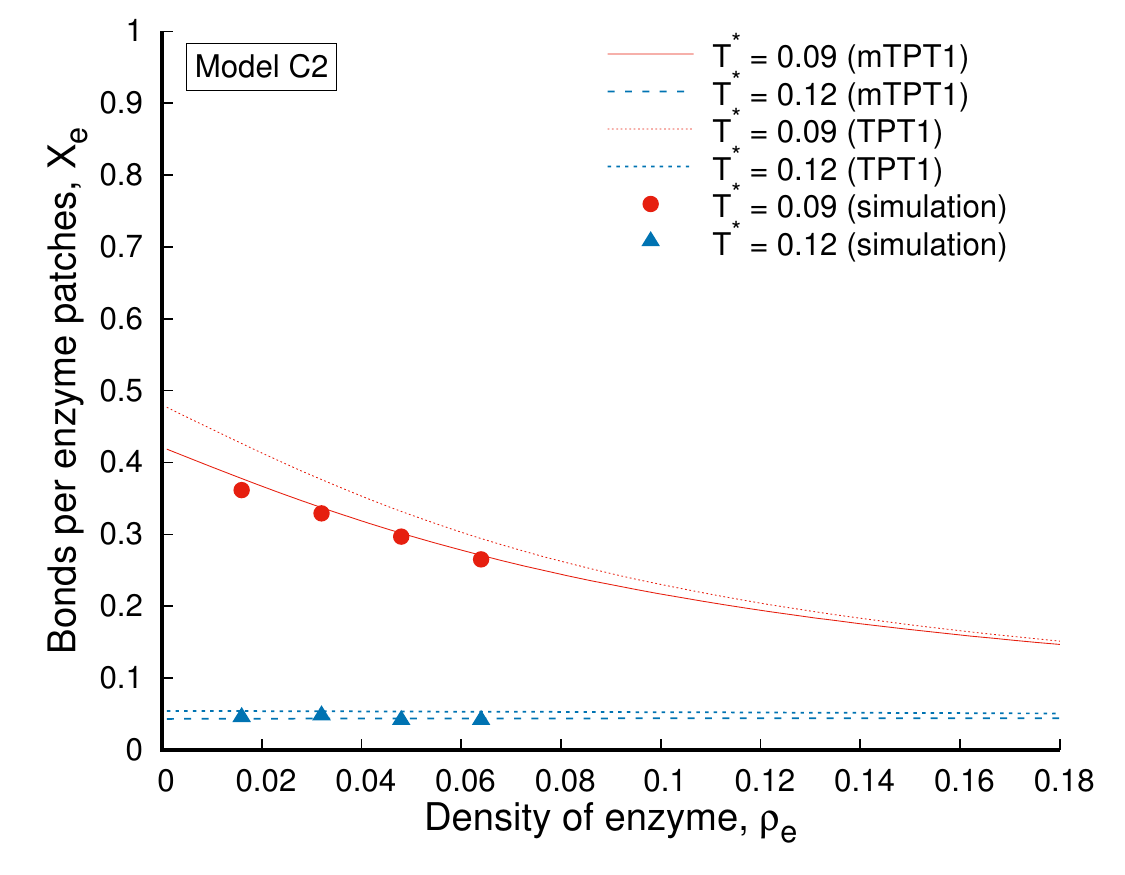}
	\includegraphics[clip,height=0.35\textwidth,angle=0]{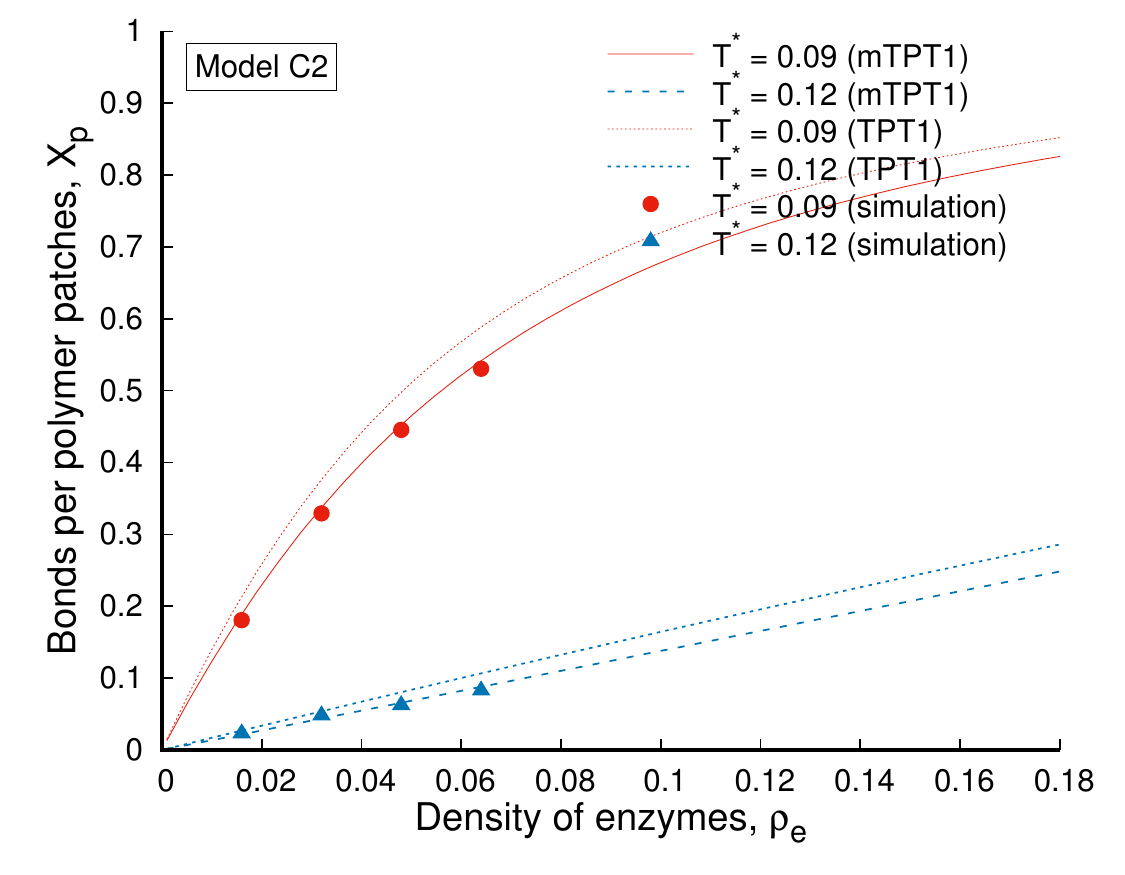}
	\caption{\label{fig:c2frac} Number of bonds per number of enzyme patches $X_e$ (left panel) and per total number of polymer functional groups $X_p$ (right panel) depending on the number density of enzyme molecules $\rho_e$ for Model~C2.}
\end{figure*}

\begin{figure*}[!h]
	\centering
	\includegraphics[clip,height=0.35\textwidth,angle=0]{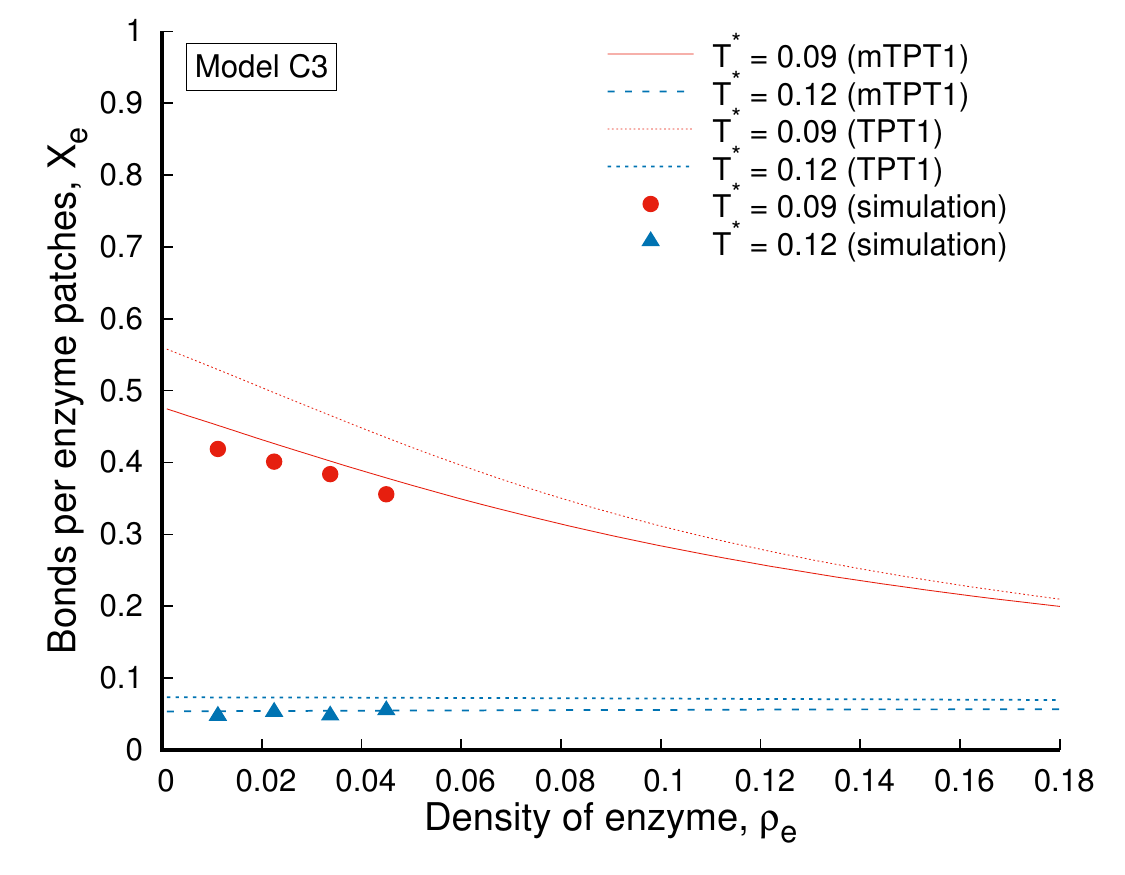}
	\includegraphics[clip,height=0.35\textwidth,angle=0]{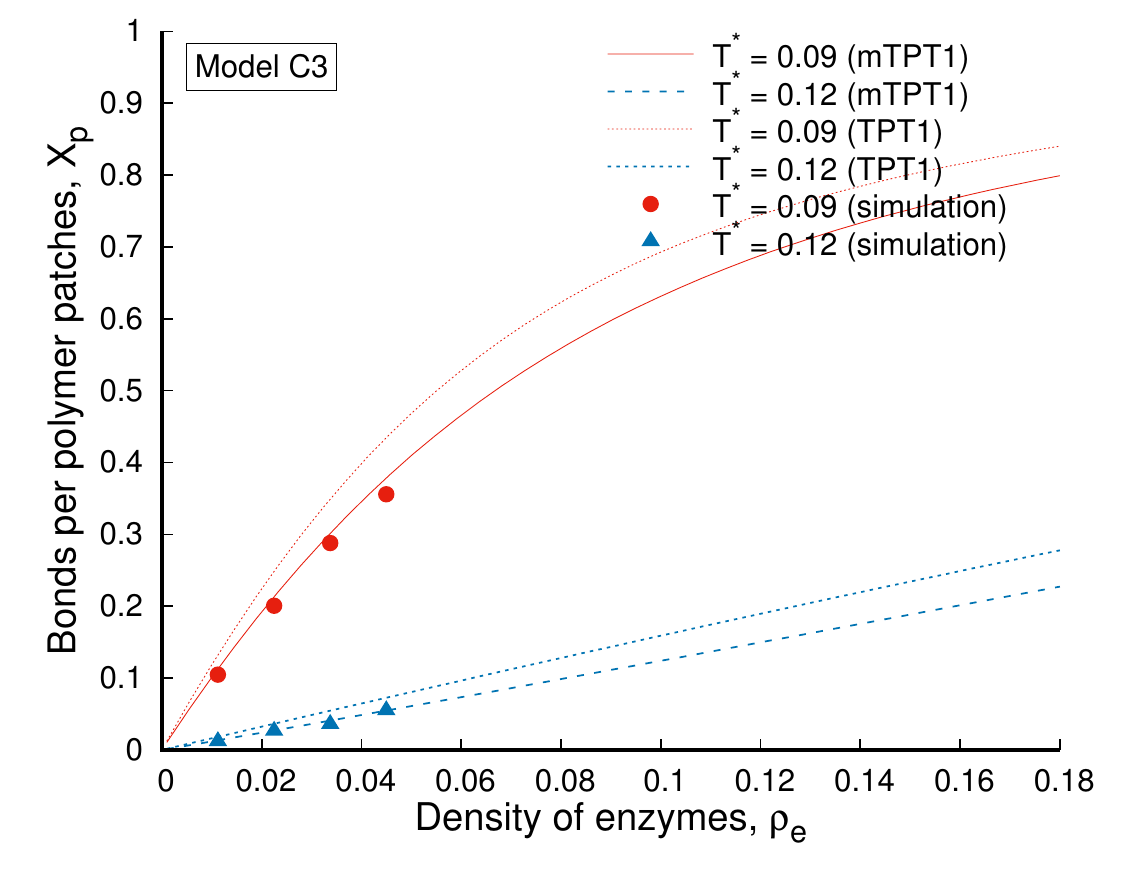}
	\caption{\label{fig:c3frac} Same as in \figurename{\ref{fig:c2frac}}, but for Model~C3.}
\end{figure*}

\begin{figure*}[!h]
	\centering
	\includegraphics[clip,height=0.35\textwidth,angle=0]{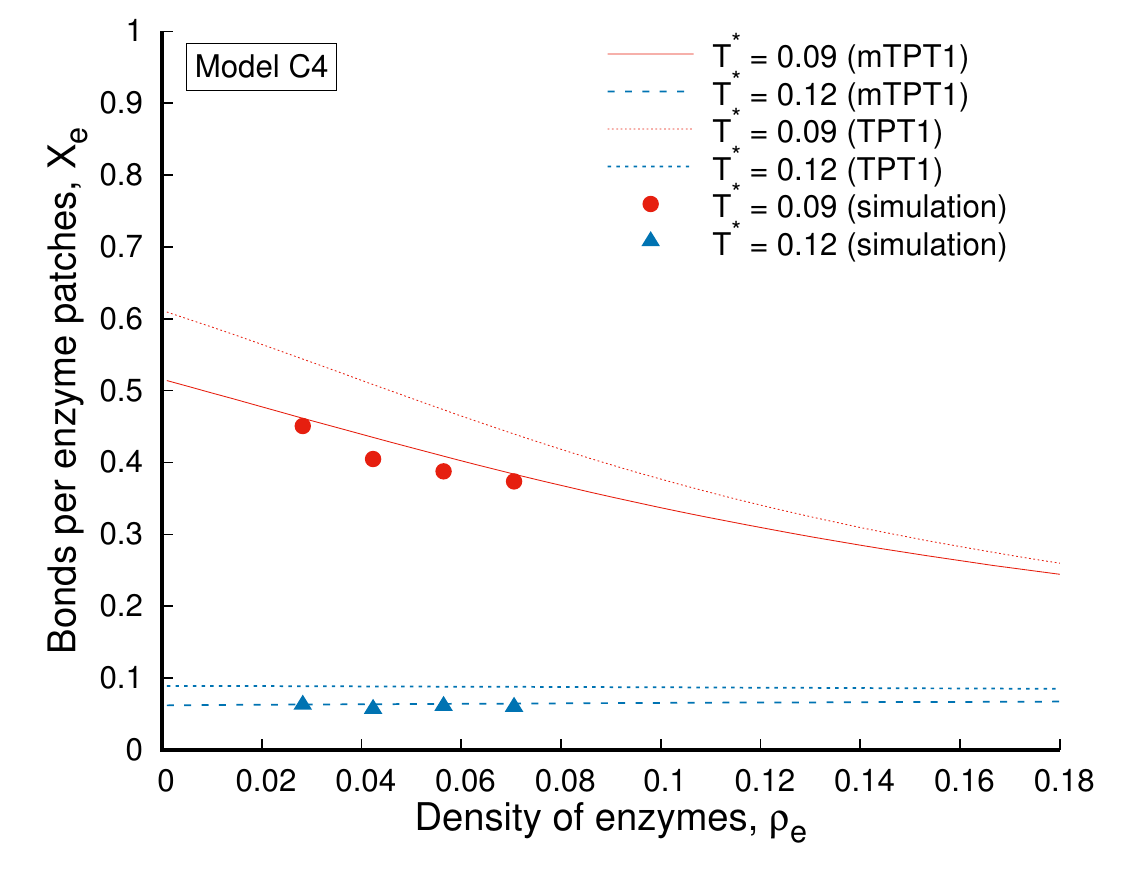}
	\includegraphics[clip,height=0.35\textwidth,angle=0]{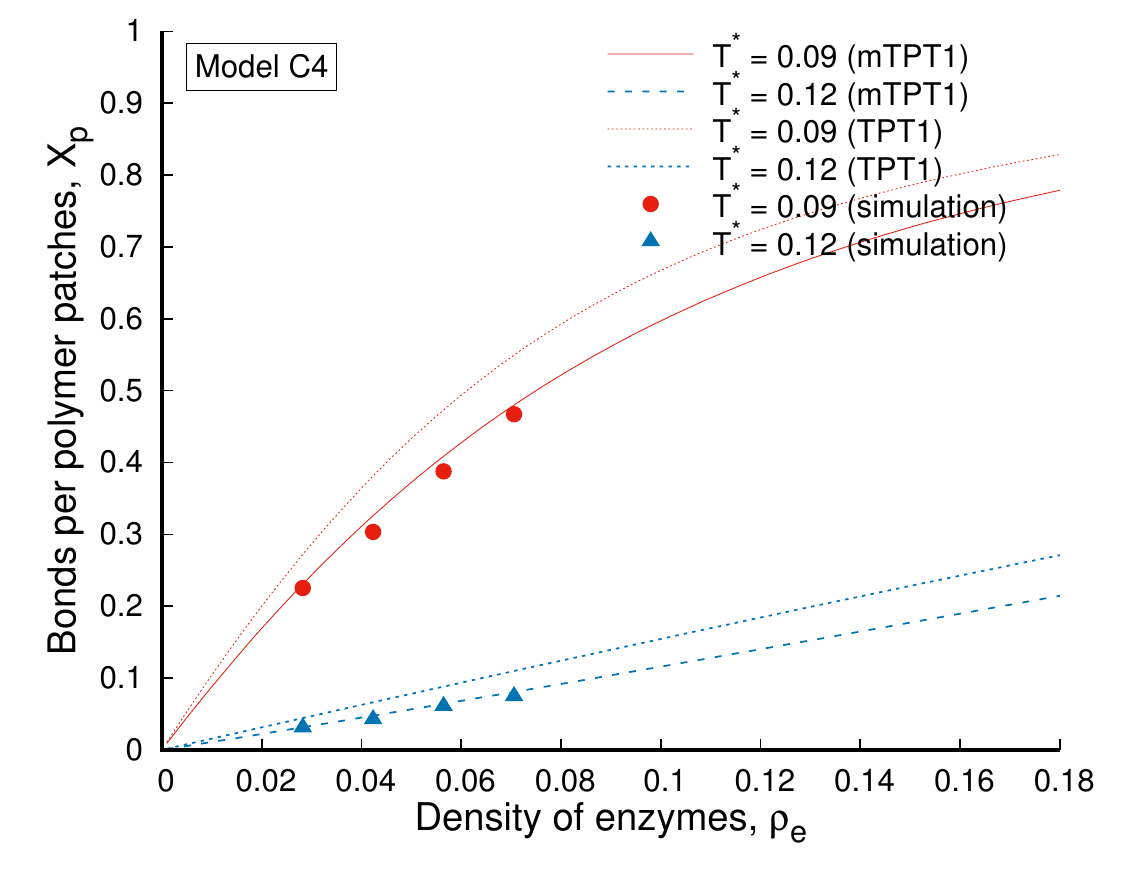}
	\caption{\label{fig:c4frac} Same as in \figurename{\ref{fig:c2frac}}, but for Model~C4.}
\end{figure*}

\begin{figure*}[!h]
	\centering
	\includegraphics[clip,height=0.35\textwidth,angle=0]{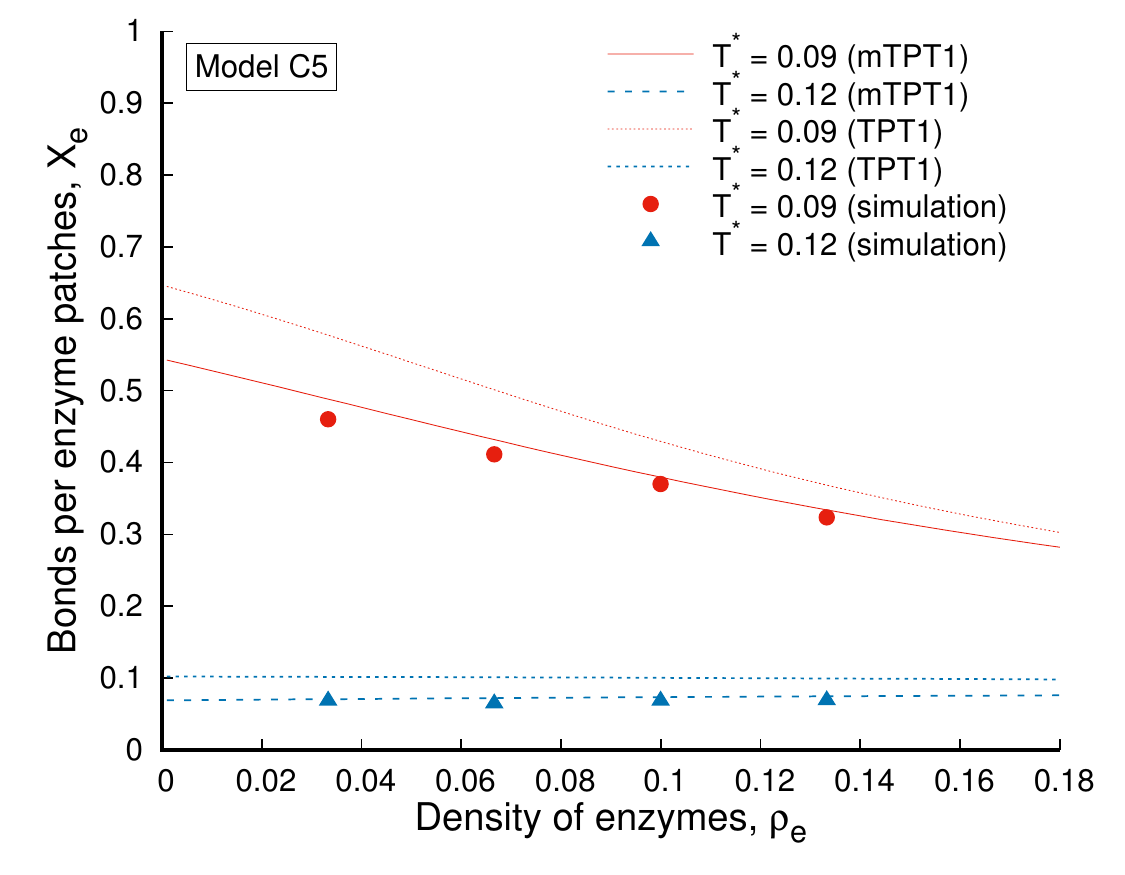}
	\includegraphics[clip,height=0.35\textwidth,angle=0]{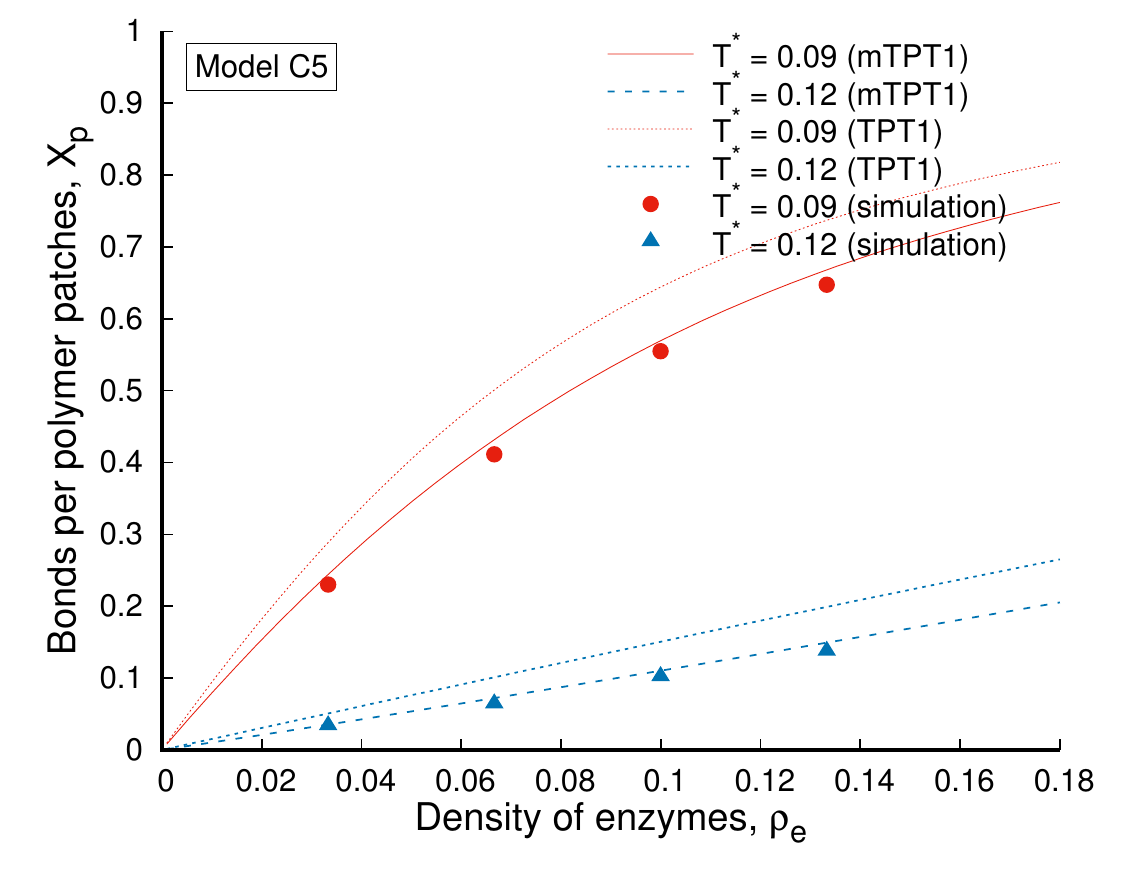}
	\caption{\label{fig:c5frac} Same as in \figurename{\ref{fig:c2frac}}, but for Model~C5.}
\end{figure*}

\begin{figure*}[!h]
	\centering
	\includegraphics[clip,height=0.35\textwidth,angle=0]{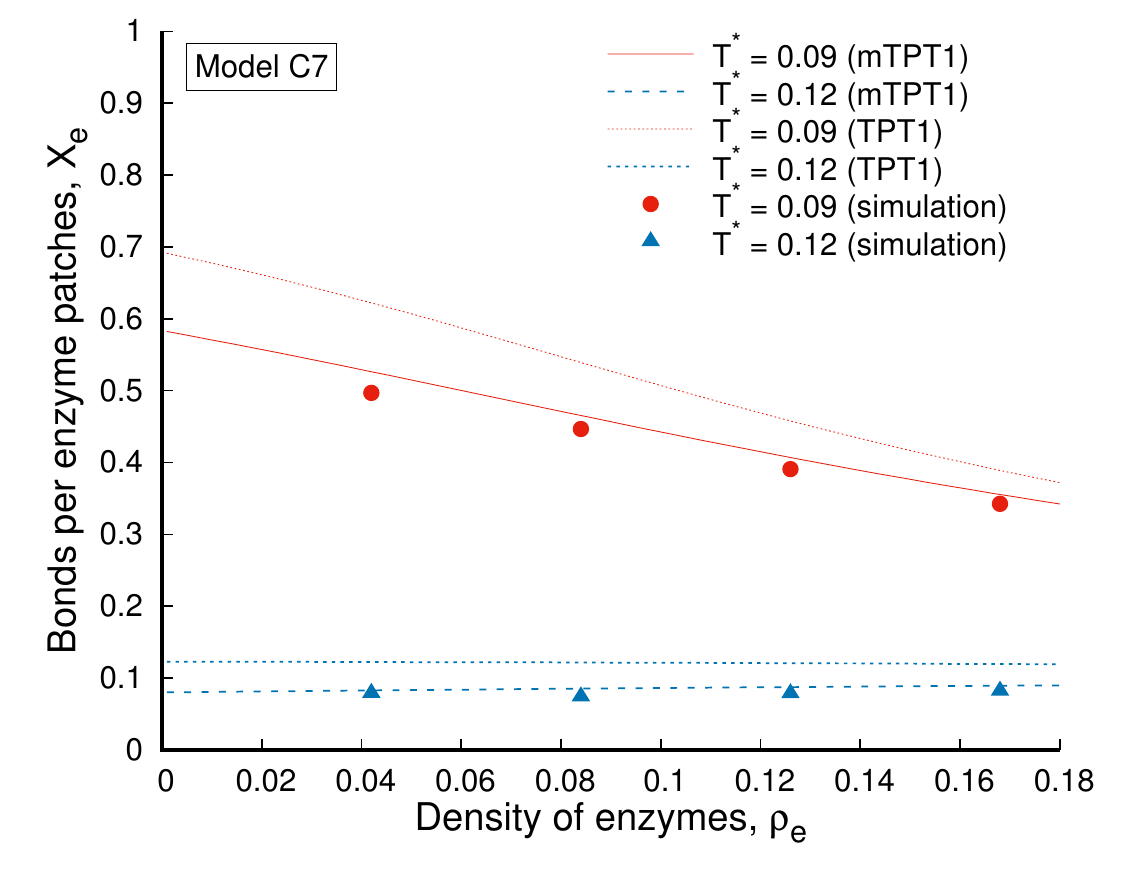}
	\includegraphics[clip,height=0.35\textwidth,angle=0]{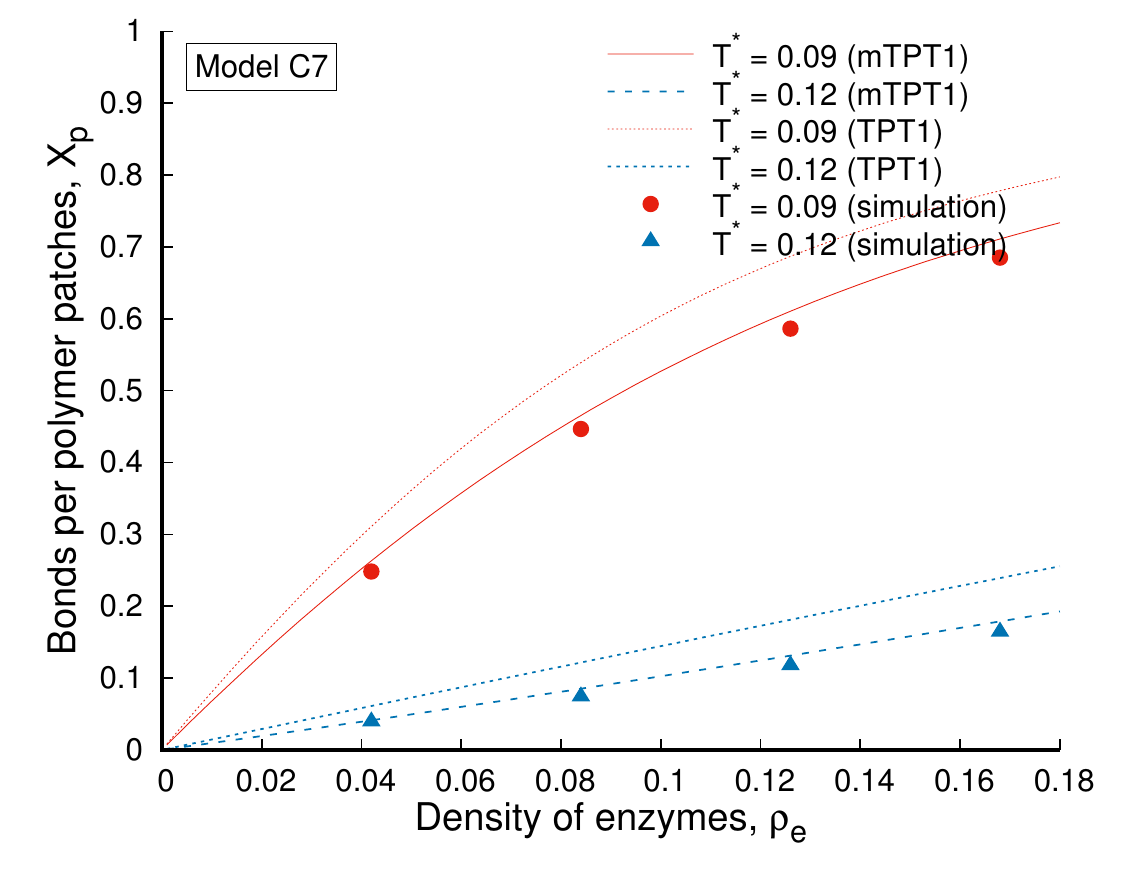}
	\caption{\label{fig:c7frac} Same as in \figurename{\ref{fig:c2frac}}, but for Model~C7.}
\end{figure*}


\begin{thebibliography}{99}

\bibitem{Wyman2011} B. Yang, Z. Dai, S.Y. Ding, and C.E. Wyman, \emph{Biofuels}, 2011, \textbf{2(4)}, 421-449.
%
\bibitem{Bao2016} G. Liu, J. Zhang, J. Bao, \emph{Bioprocess Biosyst. Eng.}, 2016, \textbf{39}, 133-140.
%
\bibitem{Chinn2015} V. Mbaneme-Smith and M.S. Chinn, \emph{Energy and Emission Control Technologies}, 2015, \textbf{3}, 23-44.
%
\bibitem{Grove2016} N.A. Carter, X. Geng and T.Z. Grove, Protein-based Engineered Nanostructures, \emph{Springer, Cham}, 2016, pp. 179-214.
%
\bibitem{Han2016} J.E. Hyeon, S.K. Shin and S.O. Han, \emph{Biotechnology J.}, 2016, \textbf{11(11)}, 1386-1396.
%
\bibitem{Bayer2010} S. Mora\"{\i}s, A. Heyman, Y. Barak, J. Caspi, D.B. Wilson, R. Lamed, O. Shoseyov and E.A. Bayer, \emph{J. Biotech}, 2010, \textbf{147(3-4)}, 205-211.
%
\bibitem{Bayer2016} J. Stern, S. Mora\"{\i}s, R. Lamed and E.A. Bayer, \emph{MBio}, 2016, \textbf{7(2)}, e00083-16.
%
\bibitem{Li2021} X. Lyu, R. Gonzalez, A. Horton and T. Li, \emph{Catalysts}, 2021, \textbf{11(10)}, 1211.
%
\bibitem{Minko2021} O. Zholobko, A. Hammed, A. Zakharchenko, N. Borodinov, I. Luzinov, B. Urbanowicz, T. Patsahan, J. Ilnytskyi, S. Minko, S.W. Pryor and A. Voronov, \emph{ACS Appl. Polym. Mater.}, 2021, \textbf{3(4)}, 1840-1853.
%
\bibitem{Bordbar2017} K. Khoshnevisan, F. Vakhshiteh, M. Barkhi, H. Baharifar, E. Poor-Akbar, N. Zari, H. Stamatis and A.K. Bordbar, \emph{Mol. Catal.}, 2017, \textbf{442}, 66-73
%
\bibitem{Minko2017} N.S. Yadavalli, N. Borodinov, C.K. Choudhury,  T. Qui\~{ñ}ones-Ruiz, A.M. Laradji, S. Tu, I.K Lednev, O. Kuksenok, I. Luzinov and S. Minko, \emph{ACS Catal.}, 2017, \textbf{7(12)}, 8675-8684.
%
\bibitem{Minko2018} X. Wang, N.S. Yadavalli, A.M. Laradji and S. Minko, \emph{Macromolecules}, 2018, \textbf{51(14)}, 5039-5047.
%
\bibitem{Beloqui2020} A. Rodriguez-Abetxuko, D. S\'{a}nchez-deAlc\'{a}zar, P. Mu\~{n}umer and A. Beloqui, \emph{Front. Bioeng. Biotechnol.}, 2020, \textbf{8}, 830.
%
\bibitem{Wertheim1984} M.S. Wertheim, {\it J.Stat.Phys.}, 1984, {\bf 35}, 19; {\it ibid.} 35.
%
\bibitem{Wertheim1986} M.S. Wertheim, {\it J.Stat.Phys.}, 1986, {\bf 42}, 459; {\it ibid.} 477.
%
\bibitem{Wertheim1987} M.S. Wertheim, {\it J.Chem.Phys.}, 1987, {\bf 87}, 7323.
%
\bibitem{Muller2001} E.A. M\"{u}ller, K.E. Gubbins, {\it Ind, Eng. Chem. Res.}, 2001, {\bf 40}, 2193.
%
\bibitem{Economou2002} I.G. Economou, {\it Ind. Eng. Chem. Res.}, 2002, {\bf 41}, 953.
%
\bibitem{Paricaud2002} P. Paricaud, A. Galindo, G. Jackson, {\it Fluid Phase Equilib.}, 2002, {\bf 194}, 87.
%
\bibitem{Tan2008} S.P. Tan, H. Adidharma, M. Radosz, {\it Ind. Eng. Chem. Res.}, 2008, {\bf 47}, 8063.
%
\bibitem{McCabe2010} C. McCabe, A. Galindo, {\it Appl. Thermodyn. Fluids}, 2010, 215.
%
\bibitem{Muller1994} E.A. Muller, L.F. Vega, K.E. Gubbins, {\it Mol. Phys.}, 1994, {\bf 83}, 1209. 
%
\bibitem{Muller1995} E.A. Muller, L.F. Vega, K.E. Gubbins, {\it Int. J. Thermophys.}, 1995, {\bf 16}, 705. 
%
\bibitem{Herdes2004} C. Herdes, J.C. Pamies, R. M. Marcos, L.F. Vega, {\it J. Chem. Phys.}, 2004, {\bf 120},
%
\bibitem{Kalyuzhnyi1994} Y.V. Kalyuzhnyi, G. Stell, M.L. Llano-Restrepo, W.G. Chapman and M.F. Holovko, \emph{J. Chem. Phys.}, 1994, {\bf 101}, 7939.
%
\bibitem{Rescic2016}, J. Rescic, Y.V. Kalyuzhnyi, P.T. Cummings, {\it J.Phys.: Cond. Matt.}, 2016, {\bf 28}, 41401.
%
\bibitem{Kalyuzhnyi2017a} Y. Kalyuzhnyi, A. Jamnic, {\it J. Mol. Liq.}, 2017, {\bf 228}, 133.
%
\bibitem{Kalyuzhnyi2017b} Y.V. Kalyuzhnyi, A. Jamnic, P.T. Cummings, {\it Soft Matter}, 2017, {\bf 13}, 1156.
%
\bibitem{Ghonasgi1994} D. Ghonasgi, W.G. Chapman, {\it J. Chem. Phys.}, 1994, {\bf 100}, 6633.
%
\bibitem{Chang1994} J. Chang, S.I. Sandler, {\it Chem.Eng.Sci.}, 1994, {\bf 49}, 2777-9822.
%
\bibitem{Kalyuzhnyi1995} Y.V. Kalyuzhnyi, P.T. Cummings, {\it J. Chem. Phys.}, 1995, {\bf 103}, 3265.
%
\bibitem{Kalyuzhnyi1997} Y.V. Kalyuzhnyi, C.T. Lin, G. Stell, {\it J. Chem. Phys.}, 1997, {\bf 106}, 1940.
%
\bibitem{Lin1998} C.T. Lin, Y.V. Kalyuzhnyi, G. Stell, {\it J. Chem. Phys.}, 1998, {\bf 108}, 6513.
%
\bibitem{Tildesley2017} M.P. Allen and D.J. Tildesley, Computer simulation of liquids, \emph{Oxford university press}, 2017.
%
\bibitem{LAMMPS} S. Plimpton, Fast Parallel Algorithms for Short-Range Molecular Dynamics, \emph{J. Comput. Phys.}, 1995, {\bf 117}, 1-19.
%
\bibitem{Espinosa2019} J.R. Espinosa, A. Garaizar, C. Vega, D. Frenkel and R. Collepardo-Guevara, \emph{J. Chem. Phys.}, 2019, 150(22), 224510.
%
\bibitem{Jover2012} J. Jover, A.J. Haslam, A. Galindo, G. Jackson and E.A. M\"{u}ller, \emph{J. Chem. Phys.}, 2012, {\bf 137(14)}, 144505.
%


\end{thebibliography}
\end{document}